\documentclass[sigconf]{acmart}

\def\BibTeX{{\rm B\kern-.05em{\sc i\kern-.025em b}\kern-.08emT\kern-.1667em\lower.7ex\hbox{E}\kern-.125emX}}

\usepackage{todonotes}

\copyrightyear{2019}
\acmYear{2019}
\acmConference{arXiv Preprint}{Sept. 2019}
\acmPrice{}
\acmDOI{}
\acmISBN{}

\renewcommand\footnotetextcopyrightpermission[1]{} % removes footnote with conference information in first column
\setcopyright{none}

\usepackage{color}
\usepackage{soul}
\usepackage{ifthen}
\begin{document}

%
% The "title" command has an optional parameter, allowing the author to define a "short title" to be used in page headers.
%\title{Pathos in Games: Exploring Techniques to Elicit Discomfort in the Audience}
\title{Pathos in Play: How Game Designers Evoke Negative Emotions}%
% The "author" command and its associated commands are used to define the authors and their affiliations.
% Of note is the shared affiliation of the first two authors, and the "authornote" and "authornotemark" commands
% used to denote shared contribution to the research.

\author{Tom Blount}

\affiliation{%
 \institution{University of Southampton}
 \city{Southampton}
 \country{UK}}
\email{T.Blount@soton.ac.uk}

\author{Callum Spawforth}

\affiliation{%
 \institution{University of Southampton}
 \city{Southampton}
 \country{UK}}
\email{C.Spawforth@soton.ac.uk}

%
% By default, the full list of authors will be used in the page headers. Often, this list is too long, and will overlap
% other information printed in the page headers. This command allows the author to define a more concise list
% of authors' names for this purpose.
%\renewcommand{\shortauthors}{Blount}

%
% The abstract is a short summary of the work to be presented in the article.
\begin{abstract}
%What is the problem?
%Digital games, as a cultural medium, are evolving to tackle a wide variety of mature and ``adult'' topics, as well as exploring themes beyond sex and violence.
%Recently, digital game designers have created games using techniques to illicit an uncomfortable emotional response in players through the use of narrative and mechanics, 
Much in the same way that people enjoy, from time to time, the pathos of consuming a tragic film or piece of literature, designers of digital games are increasingly including elements within their games that evoke uncomfortable or negative emotions in their audience, allowing players to introspect and explore ``adult'' themes and topics, including loss, regret, powerlessness, mental-health, and mortality.
%What do we do about it?
In this paper we examine a number of recent games as case studies and explore the way in which they use their mechanics to evoke feelings of discomfort in their players, and the way in which pathos serves play.
%How does this help?
Through this, we highlight a number of different techniques used by game designers and conclude by proposing further work in this space to determine exactly why players are drawn to these types of games, and to explore the ways in which research in this field could be used to drive yet more emotive and empathetic games.
\end{abstract}

%
% Keywords. The author(s) should pick words that accurately describe the work being
% presented. Separate the keywords with commas.

%
% This command processes the author and affiliation and title information and builds
% the first part of the formatted document.
\maketitle

\section{Introduction}
Many people enjoy, from time to time, the pathos from consuming a tragic film or piece of literature; the so-called ``sad film paradox'' centres around the seemingly contradictory, yet eminently observable, fact that people enjoy partaking in something that causes them to be sad~\cite{oliver1993exploring, hanich2014we}. Equally, the notion that games have to be ``fun'' to have value has long been disputed~\cite{geertz1973interpretation, jorgensen2014devil, malaby2007beyond, schechner2017performance, stenros2015playfulness}; games as a medium are able to fill a similar niche of embracing tragedy for the purpose of enjoyment and catharsis.~\citeauthor{jorgensen2016positive} describes this gratifying experience of negative emotions as ``positive discomfort''~\cite{jorgensen2016positive}.

However, as an interactive medium, games offer the unique ability to place their audience at the heart of the emotional core of their story, allowing developers to build a player's empathy for characters through the use of organic exploration and dialogue, place players in ethical quandaries that they will (hopefully) never be forced to encounter in real life, and even making the player complicit in the immoral actions of the story's characters, by enforcing their participation (rather than their passive observation) in the events of the story.

In this paper we consider a number of popular games that fulfil this role, examine the techniques the developers utilise to evoke negative feelings in their audience, and discuss the merits of these techniques, as well as areas of further research in this space. 
% \section{Background}
% Since time immemorial, media has 
% to be meaningful or entertaining. 
% 

% Similarly, the notion that games have to be ``fun'' to have value has long been disputed~\cite{geertz1973interpretation, jorgensen2014devil, malaby2007beyond, schechner2017performance, stenros2015playfulness}.
% While games as a whole have been exploring and tackling moral, social, and ethical issues almost as far back as recorded history~\cite{NEEDED} including board games and table-top role-playing games (with some Dungeon Masters being somewhat infamous for putting the party's Paladin in ethically challenging situations),

%\newpage

\section{Cases}
In this section, we explore some of the techniques used by game designers to deliberately evoke feelings of sadness or discomfort in the audience. Specifically, these are not games aimed as a means of education or therapy to survivors of loss or trauma; these are games aimed more generally at mass-market entertainment.

\subsection{Soma}
Horror games are perhaps one of the more traditional ways that audiences allow themselves to experience a negative emotional state -- fear, in this instance -- in a safe environment. \textit{Soma}~\cite{soma} is one example of a psychological horror game designed to give the player a persistent sense of dread; in this case, as they explore the underwater facility of PATHOS-II. The player is faced with a variety of experiences designed to scare them using classic film tricks such as jump-scares and disquieting audio-visual design -- for example, distorting the player's view when they look directly at a monster, which is in itself uncomfortable and helps prolong their fear of the unknown. 

This is particularly effective with combined with the interactivity inherent to games. Any action the player takes such as pushing a button, opening a door or walking forward, risks triggering one of these experiences. The result is a tense atmosphere, in which taking \emph{any} action is both uncomfortable and exciting. Notably, none of these actions allow the player to harm or delay the monsters they come across, leading the player to feel powerless against them. Reinforcing these feelings is the game's soundscape, in which the irregular clang of machinery or distant footsteps of monsters serve to remind the player of their fragility of their safety. 

% Do we talk about atmosphere/audio design/etc as techniques used to make the player uncomfortable - maybe in the context of feedback loops

\begin{figure}[H]
    \centering
    \includegraphics[width=0.9\columnwidth]{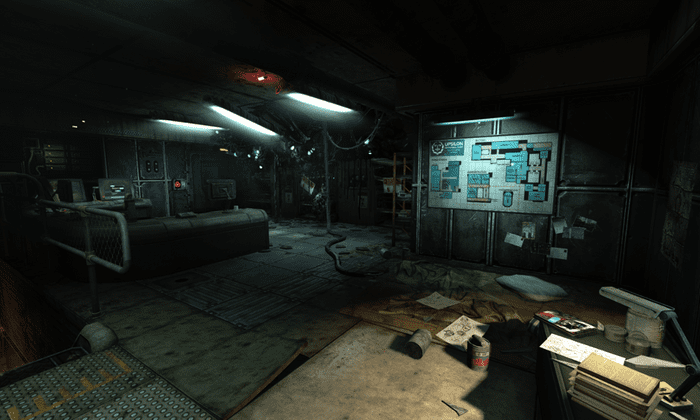}
    \caption{``Did something just move?'' \textit{Soma} discourages players from taking a close look at the monsters}
    \label{fig:examples:soma}
\end{figure}

\subsection{Depression Quest}
\textit{Depression Quest}~\citep{depressionquest} is a choose-your-own adventure game created using Twine. In the game, the protagonist suffers from  depression; the player guides them through their day to day life, in which they wrestle with dissatisfaction at work, relationship issues, and their compounding symptoms.

Throughout the game, the player is presented with a number of choices of how to act. However, as the player progresses, the more ``positive'' options (such as spending time with friends, or talking about their feelings with their partner) may become ``locked off'' due to their worsening illness. These choices can still be viewed, but are inaccessible (struck through with bold red). This serves as a metaphor for what the protagonist is experiencing, and instils a feeling of hopelessness and powerlessness in the player; they are able to see what they would like to do --- or what they ``should'' do --- but are unable to take action towards it, without having carefully and deliberately shepherded their limited resources.

\begin{figure}[H]
    \centering
    \includegraphics[width=\columnwidth]{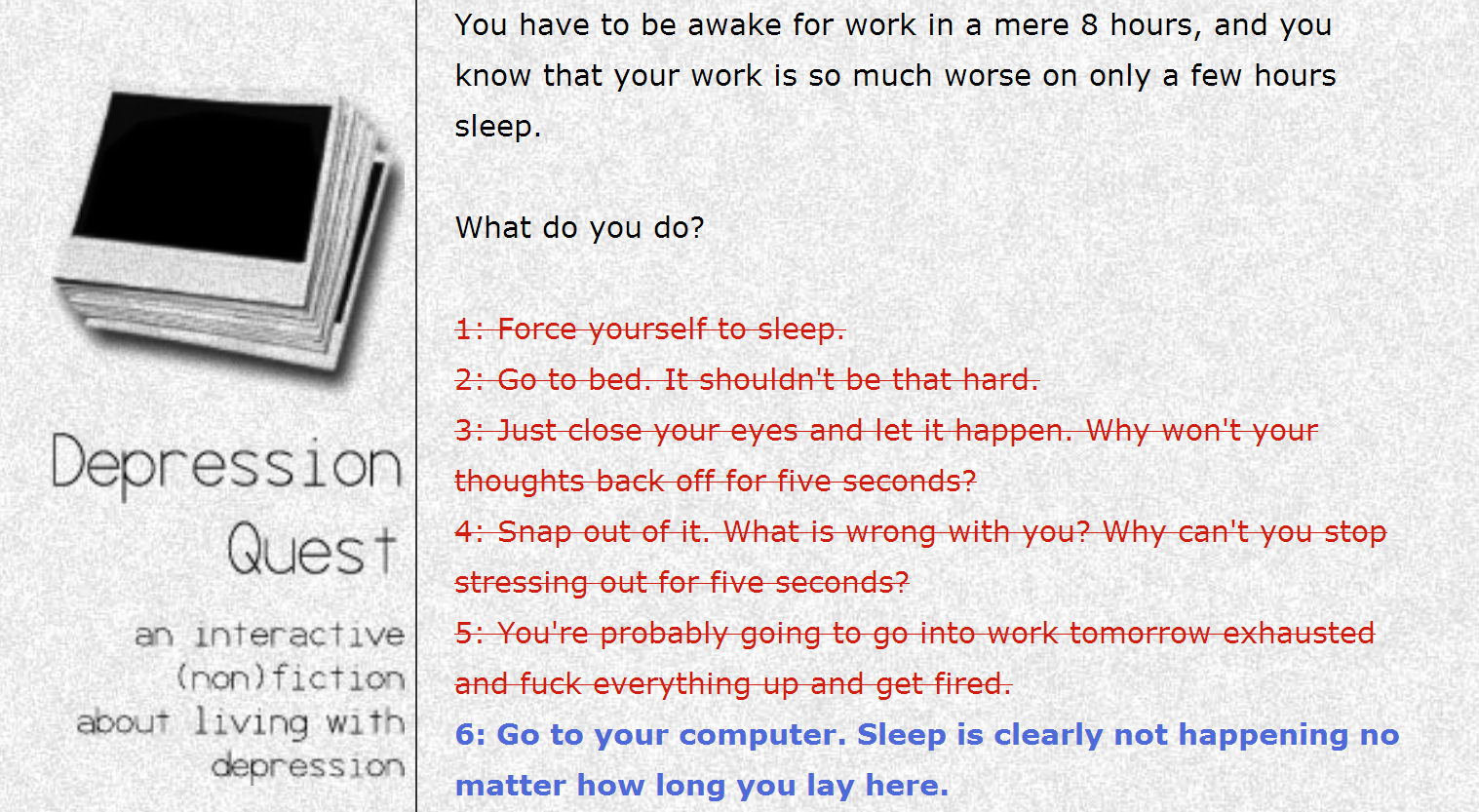}
    \caption{Some of the choices presented in \textit{Depression Quest}}
    \label{fig:examples:depression_quest}
\end{figure}

\subsection{Doki Doki Literature Club!}

\textit{Doki Doki Literature Club!}~\citep{dokidoki} is, at first glance, a typical anime-style dating-sim, in which the protagonist joins their high school's literature club in order to deepen their relationship with their classmates. However, events quickly take a subversive turn when Sayori (the protagonist's childhood sweetheart), reveals she is depressed and, shortly afterwards, hangs herself. At this point, the game restarts, with the player's previous saves erased, and Sayori notably absent from the game, the other characters having no memory of her existence.

\begin{figure}[H]
    \centering
    \includegraphics[width=0.9\columnwidth]{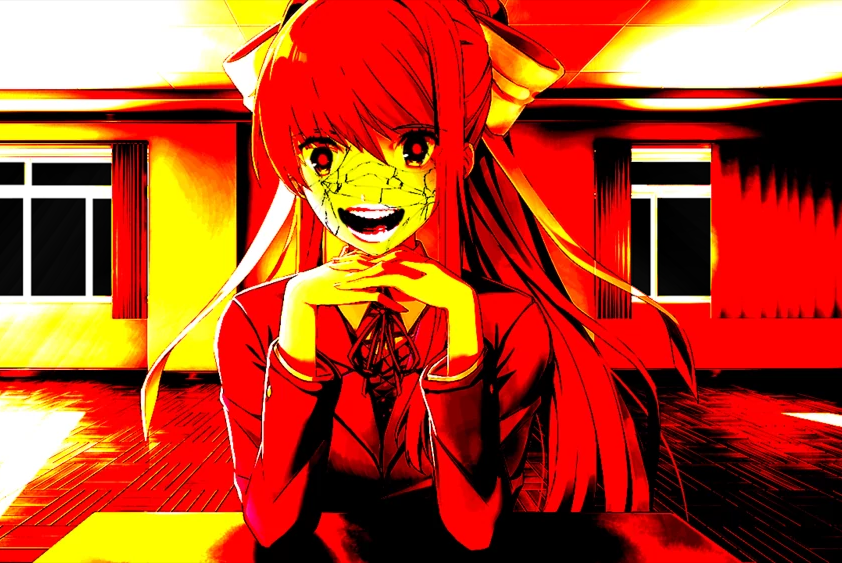}
    \caption{As the narrative of \textit{Doki Doki} progresses, reality begins to warp around the protagonist}
    \label{fig:examples:dokidoki}
\end{figure}

This pattern ultimately repeats itself, with the game appearing more and more corrupted each time (such as in Figure \ref{fig:examples:dokidoki}), until only one character, Monika, is left. In this playthrough, Monika reveals herself to be the game's ultimate antagonist who has been manipulating the files of the other characters after she became self-aware within the confines of the game. She confesses her love for the player --- not the protagonist, but the \textit{player} --- and the player is only able to progress by exiting the game and manipulating the game's files directly (mirroring the way Monika manipulated the other characters).

While this might shatter the player's immersion in one sense, in another it heightens it by creating a new narrative that acknowledges the existence of the game itself, while deliberately blurring the lines between fiction and reality.

% \subsection{Undertale}

% \textit{Undertale}~\cite{undertale} is a role-playing game in which the player takes on the role of a child that has fallen into the Underworld -- a realm of monsters hidden beneath the earth -- and must make their journey home. At that start of their adventure, the protagonist meets Toriel, a kind, motherly figure who guides them through the early stages of the game. However, later, the player must choose to kill Toriel to progress further in the game \todo[inline]{no they don't}
% Interestingly, on a second playthrough, \textit{Undertale} allows the player a chance at redemption, and it becomes possible to proceed while sparing Toriel.

% \textit{Undertale} also manages to instil a particular type of dread in its players, in a rather unorthodox fashion; at the conclusion of the game, the primary antagonist ``corrupts'' the player's game, crashing the game to the desktop, and (apparently, at least) deleting their save file. This is, of course, a bluff, and the player is given the opportunity to defeat the antagonist once and for all, and save the universe. However, it is a relatively harrowing experience the first time a player encounters it!

% \begin{figure}[H]
%     \centering
%     \includegraphics[width=0.9\columnwidth]{figures/examples/undertale}
%     \caption{To progress in \textit{Undertale}, the player must kill Toriel, their mentor and mother-figure}
%     \label{fig:examples:undertale}
% \end{figure}

\subsection{This War of Mine}
\textit{This War of Mine}~\citep{thiswarofmine} puts the player directly in control of the plight of a group of civilians caught in an ongoing civil war. The player must help them build and maintain their shelter during the day, and scavenge for resources (such as food, medicine, and building materials) during the night.

\begin{figure}[H]
    \centering
    \includegraphics[width=0.9\columnwidth]{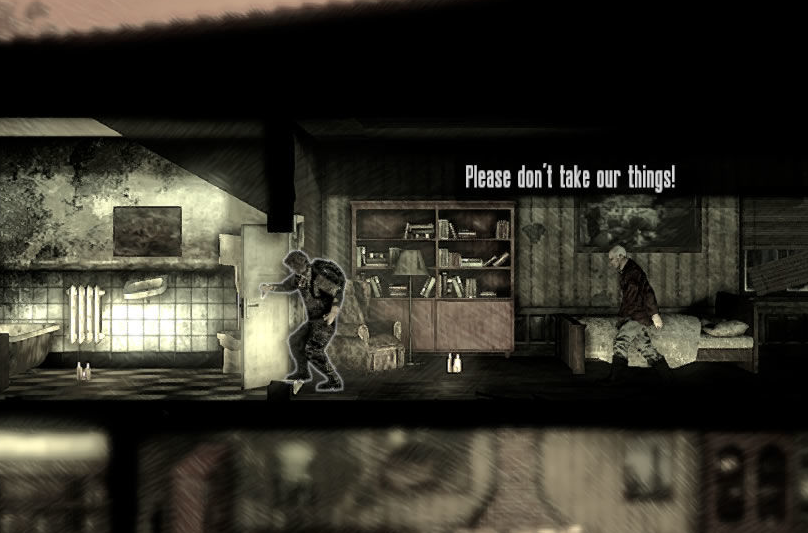}
    \caption{Stealing from an elderly couple to avoid starvation in \textit{This War of Mine}}
    \label{fig:examples:this_war_of_mine}
\end{figure}

While \textit{This War of Mine} never explicitly coerces the player into making an immoral choice, the game is rife with the opportunity, and circumstances often lead the player to take actions that make them feel uneasy. As an example, while out scavenging, the player may stumble across a soldier threatening to assault a young woman; the player is often ill-equipped to intervene, and has the choice to simply walk away and pretend they saw nothing. As another example, as the game progresses, and winter draws in, supplies become more and more scarce, subtly reinforcing more and more desperate behaviour from the player, potentially culminating in violently robbing an elderly couple for their supplies (Figure \ref{fig:examples:this_war_of_mine}), to save their own group from starvation~\citep{de2017case}.

% \subsection{The Beginner's Guide}
% \TODO{Maybe we drop this example}

% Presented as an apparent work of non-fiction, the narrator of \textit{The Beginner's Guide}~\cite{beginnersguide} has the player explore a series of (sometimes surreal) game prototypes created by his friend, ``Coda''. Over the course of the game, the narrator explains how 

% While, on the surface, the player primarily interacts with the half-finished prototypes of games, the ``goal'' of the player is more more focused on learning about, and attempting to understand, the emotional state of the narrator, and the mysterious Coda.

% At the climax of the game, it is revealed that Coda is not suffering from depression, and had previously asked the narrator to stop sharing his work (as the games were never meant for anyone else), and to refrain from contacting him again. In response, the narrator then apparently responded by releasing \textit{The Beginner's Guide} (going directly against Coda's wishes), in an attempt to regain contact.

% \TODO{Maybe this is more ``sad story'' than ``sad mechanics''; too similar to \textit{That Dragon}?}

%~\cite{thorne2017gaming}

% \begin{figure}[H]
%     \centering
%     \includegraphics[width=\columnwidth]{figures/examples/beginners_guide}
%     \caption{One of the surreal encounters in \textit{The Beginner's Guide}}
%     \label{fig:examples:beginners_guide}
% \end{figure}

\subsection{That Dragon, Cancer}
Some games are able to evoke feelings of discomfort just from their premise alone (and so much so that, while writing this paper, we had to draw straws to decide who would play this particular example). As the name suggests, \textit{That Dragon, Cancer}~\cite{thatdragoncancer} describes the true story of a couple, raising their son who was diagnosed with terminal cancer at the age of twelve months. The story is told through small vignettes of their experiences over the next four years, using point-and-click exploration and puzzle-solving. While the mechanics of this game do not inherently convey the emotional content of the game, we feel that this title merits inclusion to our case study purely down to the raw emotional force of the narrative as presented through interaction; the game encourages players to contemplate both their own mortality, and how they may have to cope with the mortality of the ones they love~\cite{schott2017dragon}.

\begin{figure}[H]
    \centering
    \includegraphics[width=0.9\columnwidth]{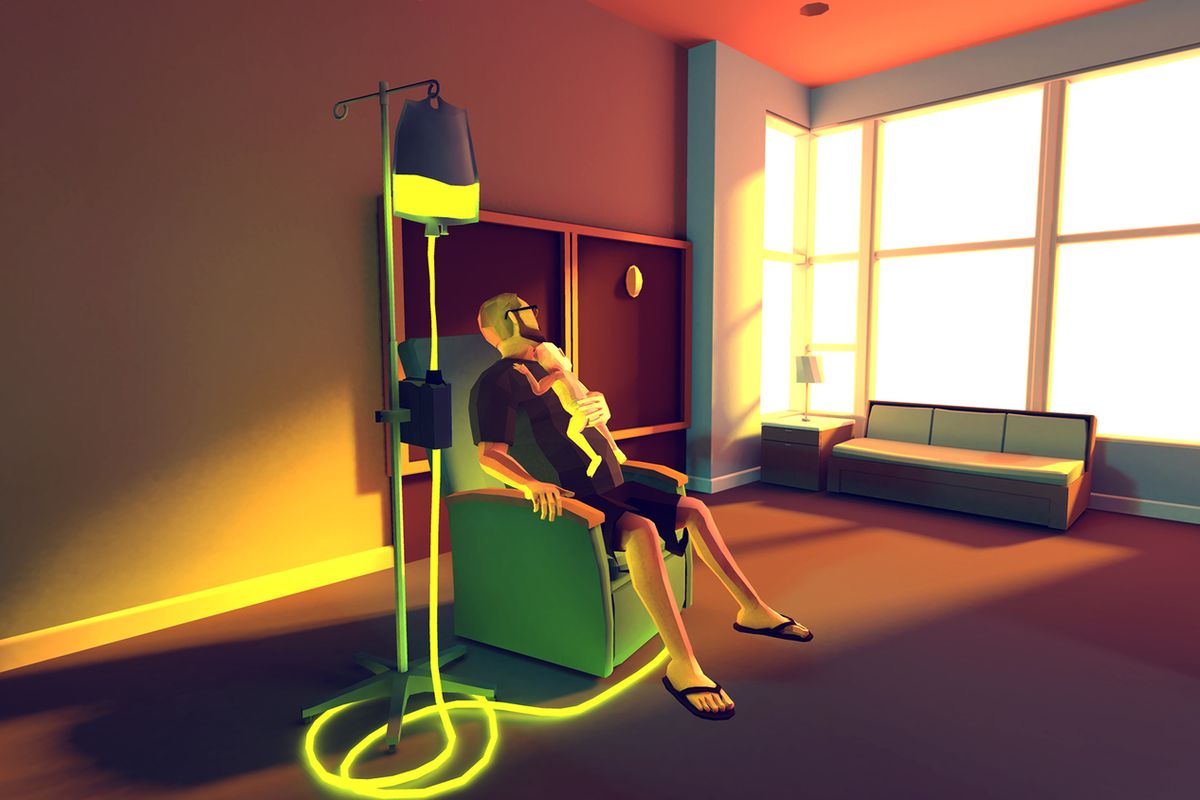}
    \caption{One of the many moving scenes captured in \textit{That Dragon, Cancer}}
    \label{fig:examples:that_dragon_cancer}
\end{figure}

\subsection{Spec Ops: The Line}

\textit{Spec Ops: The Line}~\cite{specops} has, on the surface, all the appearances of a fairly standard gung-ho, war-glorifying, third-person-shooter. However, as with \textit{Doki Doki}, this initial appearance is quickly subverted. The protagonist -- Walker, a special-forces soldier sent into a disaster-ravaged Dubai to aid the rescue of the civilian populace -- finds himself drawn into conflict with ``friendly'' American forces which continues to escalate, partly due to circumstance and mutual miscommunication, and partly due to the choices Walker makes.

\begin{figure}[H]
    \centering
    \includegraphics[width=0.9\columnwidth]{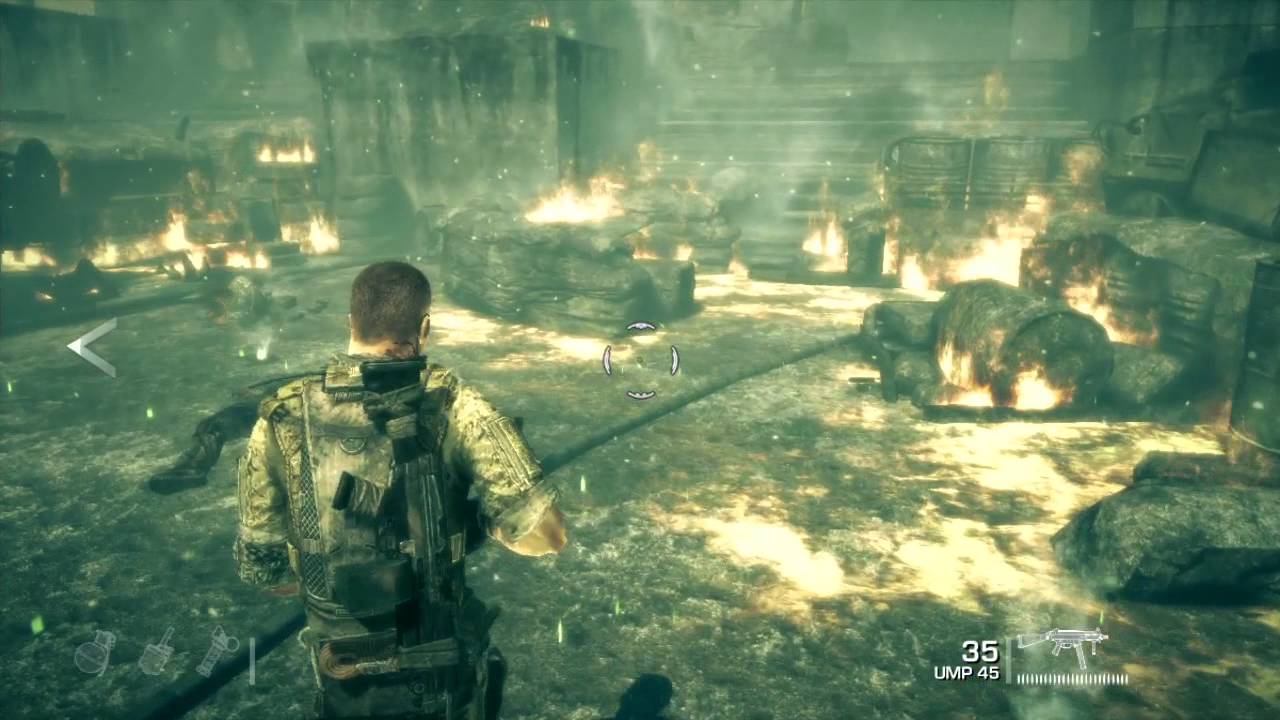}
    \caption{The aftermath of a white phosphorous attack in \textit{Spec Ops: The Line}}
    \label{fig:examples:specops}
\end{figure}

One of the pivotal moments of the game has the protagonist use white phosphorus --- a controversial incendiary weapon --- on American troops (shown in Figure~\ref{fig:examples:specops}). This is a desperate and deeply disturbing act in itself, but is worsened all the more when (unintentionally, and unavoidably) a group of refugees are caught in the crossfire. After this moment, and throughout the rest of the game, Walker begins to exhibit symptoms of post-traumatic stress disorder, suffering from hallucinations, and delusions.

%  his combat vocalisations shift from relatively stoic (\textit{``Target neutralised''}), to angry (\textit{``Got the fucker!''}), and his execution of downed enemies goes from the desperate and pragmatic to the visceral and violent.
%Despite being wholly in control of Walker's actions, the player begins to disassociate themselves from

In fact, \textit{Spec Ops} has so many of these well-designed, explicitly uncomfortable and disquieting moments of gameplay, that an entire paper could be dedicated to critiquing their meaning and impact. Fortunately, \citeauthor{keogh2013killing} has already done just that~\cite{keogh2013killing}.

Despite the more explicitly disturbing elements that the player experiences through gameplay, perhaps one of the most effective techniques to evoke a response in the player within \textit{Spec Ops} is the use, and evolution of, the game's loading screens. Initially, they contain tips for playing the game, such as how to take cover, or how to use certain weapons. However, as the game goes on, these gradually become more disturbing (\textit{``Cognitive dissonance is an uncomfortable feeling caused by holding two conflicting ideas simultaneously,''} \textit{``The US military does not condone the killing of unarmed combatants. But this isn't real, so why should you care?''}, or \textit{``To kill for yourself is murder. To kill for your government is heroic. To kill for entertainment is harmless.''}) until finally, they begin to outright blame and chastise the player for their part in the events of the game: \textit{``This is all your fault.''} \textit{``Do you feel like a hero yet?''} \textit{``If you were a better person, you wouldn't be here.''} By using influences \textit{outside} of gameplay (in conjunction with the gameplay itself) the player is directly compelled to feel uncomfortable not just at the actions of the protagonist, but at themselves for allowing those actions to happen. In a game that frequently examines the themes of choice, and choicelessness, the only real choice left to the player is to be complicit in immoral actions, or not to play at all.

\section{Analysis}
While the games presented above represent a tiny fraction of the games that evoke negative emotions in their players, we can already see a number of generalisable techniques beginning to emerge. 

\textit{Depression Quest} and \textit{Soma} both render feelings of hopelessness and powerlessness in their players by presenting the players with challenging situations, then removing the player's ability to address those challenges. These games do this in stark contrast to standard game design philosophy, in which more interactivity and choice is better; they sacrifice these in order to deliver a greater emotional blow to the player. This effect is only amplified by contrasting it with the power-fantasy prevalent in games.

%\textit{Spec Ops} and \textit{Undertale} also utilise a lack of choice as an emotional tool. However, unlike \textit{Depression Quest} and \textit{Soma}, which limit choice in order to constantly convey a particular emotional state throughout the game, \textit{Spec Ops} and \textit{Undertale} focus on a single moment, forcing the player to commit a singularly unpleasant act: killing a large number of civilians in \textit{Spec Ops}, or killing a mother-figure in \textit{Undertale}. \todo{no}
%In both cases, this is designed to make the player reflect on the actions they've just taken, despite them being presented in notably different ways. In \textit{Spec Ops}, the player only realises after they've already done the deed, whereas \textit{Undertale} presents the player with a number of options to explore, however it increasingly becomes clear that \todo[inline]{only killing Toriel will progress the game.} 

\textit{Spec Ops} and \textit{Doki Doki} share a number of similar traits in the way in which the evoke a response from their players. Firstly, both present themselves as typical examples of a particular genre, before going out of their way to subvert the established norms of that genre.
Another technique common to both \textit{Spec Ops} and \textit{Doki Doki} is breaking the fourth wall and addressing the player directly, albeit in different ways. \textit{Doki Doki} directly addresses the player, through dialogue and by ``threatening'' their save file, while \textit{Spec Ops} forces the player to reflect on their actions in the game. In both cases, the games seek to erode the wall between the safe environment of the game and reality, leaving the player more vulnerable to the game's emotional payload than they otherwise might have been.

In the case of \textit{Spec Ops}, this may have been necessary; many games (such as \textit{Grand Theft Auto}, \textit{Infamous}, or \textit{Star Wars: Knights of the Old Republic}, to name a few) allow, and even encourage, the player to not only take unambiguously ``evil'' actions, but to revel in them. These games do not evoke that same visceral level of unease that, for example, \textit{Spec Ops} creates (\citeauthor{jorgensen2016positive}'s ``positive discomfort''), instead allowing the player to experience a sense of catharsis through subverting common social taboos.

Finally, a game's premise or theming alone may be enough to evoke a sense of discomfort in the player, such as in \textit{That Dragon, Cancer}, by forcing the player to address difficult topics frequently avoided in day-to-day life. 
Other narrative-focused games such as \textit{The Beginner's Guide}, \textit{Gone Home}, or \textit{Dear Esther} use environments and exploration to convey their story and instil a sensation of melancholy and nostalgia in their players.
This can be compared (and contrasted) with games which use deliberately shocking material and imagery that do so to generate outrage in the popular eye as a means of marketing, rather than providing the player with any sort of deeper emotional experience.
%A similar example might be seen in the game \textit{Rape Day}, a rape simulation game banned from the games distribution platform Steam over its content. Interestingly, this is the only category of games that we've been highly reluctant to play (and haven't, in the case of the latter game).

%Overall,

\section{Conclusions and Future Work}

% \begin{itemize}
%     \item Determining WHY people play sad games? *Is* it the same reason that they watch sad films? 
%     \item What about games that "force" the player to do bad things? 
%     \item Examining the effect of decision making in 1st vs 3rd person
%     \item Same with regard to empathy
%     \item Extending both of these to include Virtual Reality (does violence in VR can feel more visceral/disquieting? What about suicide in VR?)
%     \item Games that requite the \textit{player} to take ethically dubious actions (e.g. trespassing in StoryPlaces)
% \end{itemize}
In this paper, we have examined a number of techniques used by game designers, such as manipulating the player's sense of perception, limiting the choices available to the player, encouraging (and outright forcing) them to commit actions that they would otherwise choose not to, and ``meta-game'' elements which toy with the accepted norms of games, such as breaking the fourth wall or manipulating loading screens and save files.

Our intention is that this discussion inspires new and interesting avenues of research. For example, although we discuss several different techniques used by designers, in this work we only discuss our own anecdotal experiences of how these games affected us. A broader study could aim to do further this work twofold: firstly, to construct a complete taxonomy of the design techniques used to evoke negative emotions in games, and secondly to thoroughly deconstruct the range of players' emotional responses to these techniques. In addition, it may be beneficial to seek a better understanding of what draws players to these games in general, and whether there is an overlap with similar media such as films or literature.

Another avenue in this direction of research is exploring how the level of immersion~\cite{brown2004grounded} affects the level of emotional impact, both in terms of empathy with non-player characters, and with the willingness (and ability) to perform particular acts. For example, are players more impacted when witnessing events from a first-person perspective, than they are from a third-person perspective (despite maintaining the same ability to interact with the virtual world around them). This could be further extended to investigate virtual reality, and they ways in which players react to some of the techniques described above -- as well as more ``traditional'' actions encountered in games, such as the application of violence -- in a wholly immersive environment. Taking this to the extreme, this type of research could be extended further to encapsulate the genre of ``Alternate Reality Games''~\cite{kim2009storytelling} in which the lines between game and reality are blurred even further. However, the findings from work such as this may not be applicable to more traditional games media.

Overall, we aim to reiterate that games do not have to be ``fun'' to be absorbing or engaging; there is clearly a desire for a wide range of games that allow players -- or give players the opportunity to allow themselves -- to explore an equally wide range of emotions, both positive and negative.

%
% The acknowledgments section is defined using the "acks" environment (and NOT an unnumbered section). This ensures
% the proper identification of the section in the article metadata, and the consistent spelling of the heading.
% \begin{acks}

% \end{acks}

%
% The next two lines define the bibliography style to be used, and the bibliography file.
\bibliographystyle{ACM-Reference-Format}
\bibliography{bibliography}

\end{document}